\documentclass[a4paper,fleqn]{cas-dc}

\usepackage[numbers,sort&compress]{natbib}

\usepackage{epstopdf}
\usepackage{lineno,hyperref}
\usepackage{amsmath,amssymb,amsfonts}
\usepackage{algorithmic}
\usepackage{graphicx}
\usepackage{textcomp} 

\usepackage[figuresright]{rotating}
\usepackage{algorithm}  
\usepackage{algorithmic} 

\usepackage{bm}
\usepackage{float}
\usepackage{subfigure}

\usepackage{multirow,threeparttable}   
\usepackage{fancybox}

\begin{document}

\let\WriteBookmarks\relax
\def\floatpagepagefraction{1}
\def\textpagefraction{.001}

\shorttitle{LRBmat}    



\title[mode = title]{
	LRBmat: A Novel Gut Microbial Interaction and Individual Heterogeneity Inference Method for Colorectal Cancer
}




\author[1]{Shan Tang}
\fnmark[1]
\author[1]{Shanjun Mao\corref{cores}}
\fnmark[1]
\cormark[1] 
\ead{shjmao@hnu.edu.cn}
\author[2]{Yangyang Chen}
\fnmark[1]
\author[1]{Falong Tan}
\author[3]{Lihua Duan}
\author[4]{Cong Pian}
\author[5]{Xiangxiang Zeng}

\address[1]{Department of Statistics, Hunan University, Changsha 410006, China}
\address[2]{Department of Computer Science, University of Tsukuba, Tsukuba 3058577, Japan}
\address[3]{Department of Rheumatology and Clinical Immunology, Jiangxi Provincial People's Hospital, The First Affliated Hospital of Nanchang University, Nanchang 330006, China}
\address[4]{College of Sciences, Nanjing Agricultural University, Nanjing 210095, China}
\address[5]{Department of Computer Science, Hunan University, Changsha 410086, China}

\fntext[1]{These authors contributed equally to this work.}
\cortext[cores]{Corresponding authors: Shanjun Mao.}

















\begin{abstract}
    Many diseases are considered to be closely related to the changes in the gut microbial community, including colorectal cancer (CRC), which is one of the most common cancers in the world. The diagnostic classification and etiological analysis of CRC are two critical issues worthy of attention. Many methods adopt gut microbiota to solve it, but few of them simultaneously take into account the complex interactions and individual heterogeneity of gut microbiota, which are two common and important issues in genetics and intestinal microbiology, especially in high-dimensional cases. In this paper, a novel method with a Binary matrix based on Logistic Regression (LRBmat) is proposed to deal with the above problem. The binary matrix can directly weakened or avoided the influence of heterogeneity, and also contain the information about gut microbial interactions with any order. Moreover, LRBmat has a powerful generalization, it can combine with any machine learning method and enhance them. The real data analysis on CRC validates the proposed method, which has the best classification performance compared with the state-of-the-art. Furthermore, the association rules extracted from the binary matrix of the real data align well with the biological properties and existing literatures, which are helpful for the etiological analysis of CRC. The source codes for LRBmat are available at \url{https://github.com/tsnm1/LRBmat}.
\end{abstract}



\begin{highlights}
  \item The proposed binary matrix can better simultaneously solve the problem of individual heterogeneity and gut microbial interactions, without crossterms.
  \item The proposed LRBmat is a simple and computationally inexpensive algorithm for accurately estimating the binary matrix.
  \item Although LRBmat is based on Logistic model, it can be directly extended to other common classification models, and basically have a certain improvement effect.
  \item The binary matrix can be used to detect the complex combined effect of multiple gut microbes on CRC, through association rule algorithm. 
\end{highlights}

\maketitle

\section{Introduction}
\label{sec:introduction}

Colorectal cancer (CRC) is one of the most common cancers in the world, ranking third in incidence and second in mortality, with more than 935,000 deaths each year \cite{sung2021global,biller2021diagnosis,xie2020comprehensive}. More and more researchers are focusing on CRC, and a more effective understanding of the biological property and diagnostic classification of CRC patients is a key requirement for the implementation of individualized treatment \cite{joanito2022single,wang2019molecular}. However, CRC is largely an asymptomatic disease until it reaches an advanced stage. Its diagnosis is usually through colonoscopy. Although colonoscopy is relatively simpler to identify advanced lesions, it is much difficult to detect the early-stage \cite{DEKKER20191467,wang2022afp}. There is a lot of evidence to suggest that certain components of the gut microbiota are associated with the occurrence and progression of CRC. For example, the Peptostreptococcus anaerobius (\emph{P. anaerobius}) and Fusobacterium nucleatum (\emph{F. nucleatum}) are enriched in the faecal samples and mucosal microbiota of patients with CRC \cite{long2019peptostreptococcus,hashemi2019fusobacterium}. Similarly, gut microorganisms such as Streptococcus bovis (\emph{S. bovis}), Enterotoxigenic Bacteroides fragilis (ETBF), Enterococcus faecalis (\emph{E. faecalis}) and Escherichia coli (\emph{E. coli}) are also considered to be closely related to CRC \cite{alhinai2019role,cheng2020intestinal,janney2020host}. Furthermore, the alterations in gut microflora diversity are associated with CRC development, such as ecological dysbiosis of the gut microbiota \cite{somathilaka2022graph,ahmed2018microbiome,zou2018dysbiosis}. Therefore, the aetiology of CRC can not be attributed to the presence or activity of a single microorganism, and the role of the microbial community should also be considered \cite{louis2014gut}. For instance, stool profiling shows that the gut microbial composition are differentially represented in patients with CRC compared to healthy controls (HC), and there are also differences in the diversity of microbial community structure at specific gut sites during the development of CRC \cite{weir2013stool,zhang2021differential}. These differences contribute to developing the early detection and feasible treatments for CRC, and thus improve the patient's survival. Hence, it's reasonable and viable to adopt the gut microbial data for the diagnostic classification of CRC.

There are many challenges in using gut microbial communities for CRC classification. One of the most common is the inherent characteristics of microbiota, of which microbial interaction \cite{xu2022caconet,maity2011powerful,weersma2020interaction} and individual heterogeneity \cite{liu2021microbial,chen2016two} are the two most important. The study of variable interactions is an essential issue in biology, since the gut microbiome itself is an ecosystem that involves complex interactions. In the past, the role of microbiota in human has been studied mainly through differences in microbial abundance, but the effect of microbial interactions in similar contexts has been less considered \cite{chen2020gut,krautkramer2021gut}. On the other hand, many recent studies have begun to pay more attention to heterogeneity \cite{fletcher2007heterogeneity,bedard2013tumour,nunes2020definition}. In term of CRC, heterogeneity means that there are individual differences in the same gut microbes. For example, \cite{wang2019lactobacillus} discussed that the treatment with bacteriocin-producing or non-bacteriocin-producing lactic acid bacteria (LAB) strains would have different effects on the host. \cite{jia2020metagenomic} pointed out that the bile acid metabolism of the gut microbiota would have beneficial and harmful effects on the health of different hosts. Moreover, there are many connections between the two issues. When examining whether an effect is modified by a variable or examining the heterogeneity of the observed effect across subsets of individuals, such as treatment benefit in clinical trials, the statistical term for heterogeneity is exactly `interaction' \cite{altman1996statistics}. \cite{kontopantelis2018investigating} also showed that the effect of heterogeneity could be studied in the model with interaction terms. Consequently, it is of significance to consider gut microbiota interactions and individual heterogeneity in the diagnostic classification of CRC, and it is also feasible to consider both issues at the same time.

There are many methods to diagnose cancer using gut microbes and machine learning is the most popular one, such as Logistic model \cite{cakmak2021classification}, Random Forests (RF) \cite{eck2017robust}, Support Vector Machine (SVM) \cite{zheng2020specific} and XGBoost \cite{he2021preoperative}. \cite{bang2019establishment} proposed a multi-classification model, which used machine learning to classify six diseases, including CRC, and the highest performance was observed when using gut microbial data at the genus level. When studying the microbial interaction in CRC classification, the idea of exhaustive method or kernel machine method is usually added to the above machine model \cite{bertsimas2017logistic,larsen2000interpreting,cho2022nonlinear}. Considering the example of the logistic model, which is one of the most popular classification methods and has good interpretability and prediction effect. Adding interaction terms or kernels capturing interactions to the logistic model can effectively solve the inter-relationships between gut microbes \cite{Dingen2018regressionexplorer}. However, the kernel machine approach often has relatively high computational complexity, and exhaustive methods can not be extended to search for higher-order interactions, because the number of interaction terms increases exponentially with the increase of interaction order \cite{rainey2016compression,cordell2009detecting}. Therefore, it is worthy of study to effectively take into account the interactions of gut microbes without adding too much burden, i.e., interaction terms. Although the above methods maybe perform well when there is complex interactions between gut microbes, they rarely consider individual heterogeneity. In the usual model setting, the effect of one covariable on outcome may be measured by a fixed parameter, but in many common cases, this kind of effect may vary with different individuals. In general, some interaction terms of grouped variables and covariables are considered to be added into the model to deal with heterogeneity \cite{zhao2012estimating}. However, this may not be appropriate when the covariable such as gut microbiota is continuous or more than one covariable has individual heterogeneity problem, and its effect is not always good. Moreover, it is also difficult to deal with the high dimensional situation. Thus, a novel method, without any redundant cross terms, for simultaneously considering gut microbial interaction and individual heterogeneity is needed.

\begin{figure*}[ht]
	\centering
	\includegraphics[scale=0.6]{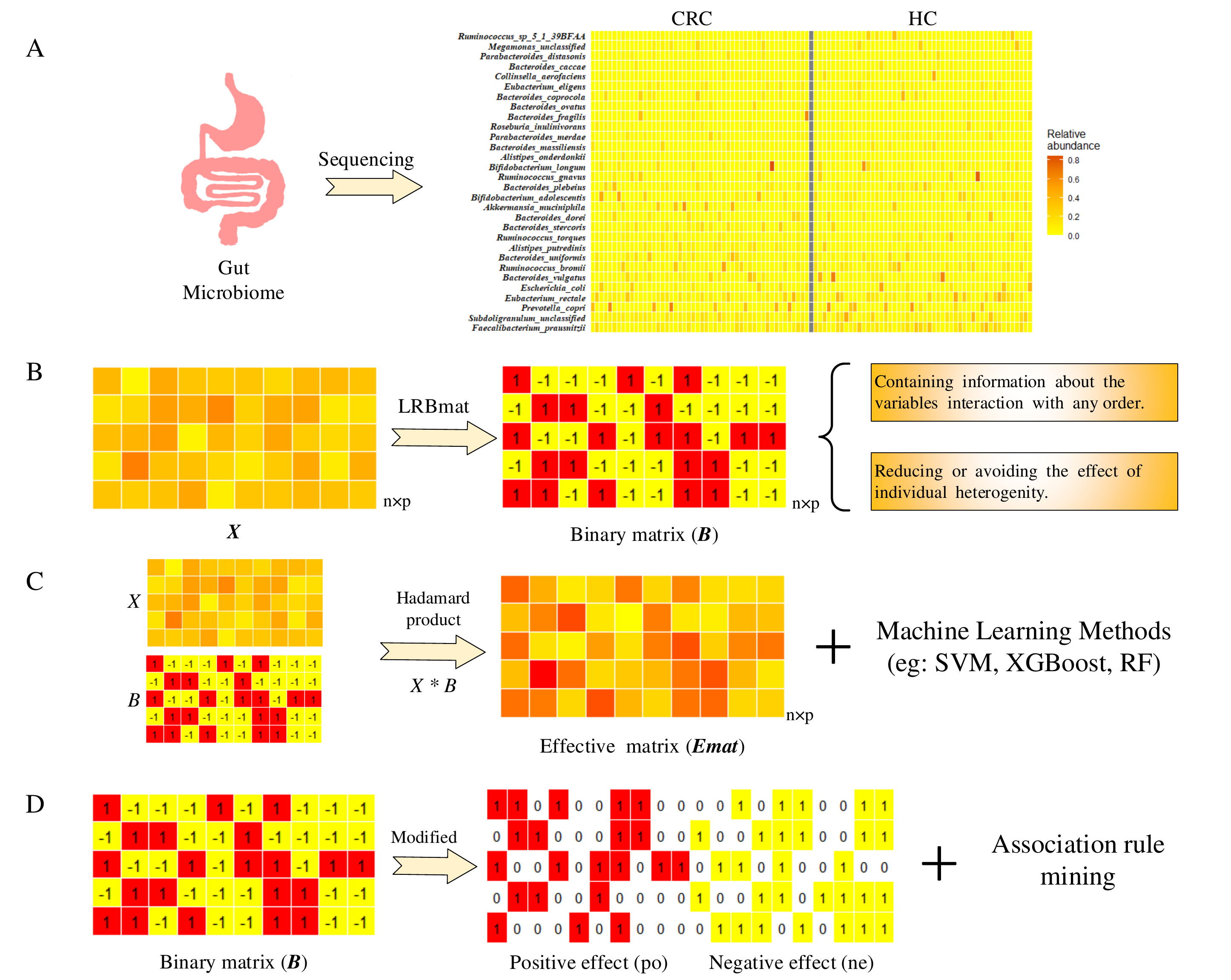}
	\caption{The Architecture of LRBmat. (A) For a sequencing example, relative abundance of the top 30 gut microbiota genomes from 100 subjects in the CRC (n = 50) and HC (n = 50) groups. (B) The binary matrix is estimated from the data through LRBmat. $\bm{X}$ is the local presentation of intestinal microbial data in (A). (C) The binary matrix can modify data $\bm{X}$ via Hadamard product, and then the produced effective matrix $Emat$ can be combined with common machine learning methods to obtain better performances. (D) After simple modification of the binary matrix, the association rule mining algorithm can be directly used to mine valuable rules.}
	\label{fig_lrbmat}
\end{figure*}

In this paper, in order to avoid the problem of excessive interaction terms of gut microbiota and effectively deal with the individual heterogeneity of microbes simultaneously, a binary matrix containing only $1$ and $\text{-}1$ is introduced, which can represent interacted effects of different gut microbes on different individuals. This is inspired by the idea that the positive or negative regulation of gene regulator \cite{grah2020relation, escoter2019fighting} and binary treatment indicators in randomized clinical trials \cite{tian2014simple,callegaro2017testing}. A simple method called LRBmat (Binary matrix based on Logistic Regression, Figure~\ref{fig_lrbmat}) is proposed to get a good estimate of this matrix. Moreover, an intuitive interpretation of the binary matrix is introduced to newly define the relationship between variable interaction and individual heterogeneity. In addition, based on the binary matrix, the association rule mining algorithm \cite{tandon2016inferring,chyou2021identifying} can be adopted to study some reasonable and valuable association rules between multiple gut microbes and CRC diseases in a subset of gut microbes. The remaining content of the paper is organized as follows: the proposed model is presented in Section~\ref{s:model}, including the binary matrix and the detailed algorithm; main results about simulations and the real data analysis are demonstrated in Section~\ref{s:result}; Section~\ref{s:conclusion} concludes the paper.

\section{Materials and Methods}

\label{s:model}

\subsection{Gut microbiota data in colorectal cancer}

To effectively illustrate the superiority of the proposed method in CRC, public metagenomic data from seven geographically diverse and pre-processed gut microbes are collected in this paper. Table~\ref{tab:sources1} demonstrates the size and characteristics of the seven datasets. There's a total of 1224 individuals, including 605 patients with CRC and 619 HC, comprising 845 gut microbial species.

Notation: the gut microbial data are expressed as $\bm{X} = (x_{i,v})_{n \times p}$, where $n$ and $p$ is the total number of individuals and gut microorganisms, respectively, $x_{i,v}$ represents the count of the $v$-th gut microorganism in the $i$-th individual and satisfies $\sum_{v=1}^{p} x_{i, v}=100$. Let $\bm{y}=\left(y_{1}, y_{2}, \cdots, y_{n}\right)^{T},\;y_{i}=0\; \text{or}\; 1 $ denote the diagnostic labels of $n$ individuals, with 0 representing HC and 1 representing patients with CRC. Thus, gut microbes and the diagnostic label are covariates and response variable, respectively.

\begin{table*}[htbp]  
    \centering
    \caption{Size and characteristics of the seven datasets.}
    \begin{threeparttable}
        \begin{tabular}{@{}llccc@{}}
            \hline
             {BioProject (ref.)}                                                 & {Group (n)}       & {Age $(average \pm s.d.)$} & {Sex $F(\%)/M(\%)$} & {Country} \\ \hline
            \multirow{2}{*}{\parbox[l]{3.2cm}{PRJEB6070  \cite{zeller2014potential}}}  & CRC (53)          & {$66.8 \pm  10.9$}       & {$45.2/54.8$}       & \multirow{2}{*}{{France}}  \\
                                                                                & HC (88)\tnote{a}  & {$60.6 \pm 11.4$}      & {$54.1/45.9$}              &           \\
            \multirow{2}{*}{\parbox[l]{3.2cm}{PRJEB7774 \cite{yang2019enterotype}} } & CRC (46)           & {$67.1\pm 10.9$}         & {$39.1/60.9$}        & \multirow{2}{*}{{Austria}}   \\
                                                                                & HC (63)\tnote{a}  & {$67.1\pm 8.5$}            & {$41.3/58.7$}      &           \\
            \multirow{2}{*}{\parbox[l]{3.3cm}{PRJEB12449  \cite{vogtmann2016colorectal}}} & CRC (52)             & {$61.8 \pm 13.6$}       & {$28.8/71.2$}         & \multirow{2}{*}{{USA}} \\
                                                                                & HC (52)           & {$61.2\pm 11$}             & {$28.8/71.2$}      &           \\
            \multirow{2}{*}{\parbox[l]{3.2cm}{PRJNA447983  \cite{thomas2019metagenomic}} } & CRC (61)          & {$64.6\pm 8.3$}         & {$24.6/75.4$}         & \multirow{2}{*}{{Italy}}   \\
                                                                                & HC (52)           & {$62.5\pm 7.8$}            & {$44.2/55.6$}      &           \\
            \multirow{2}{*}{\parbox[l]{3.2cm}{PRJEB27928  \cite{wirbel2019meta}} }  & CRC (60)\tnote{a}     & {$63.5\pm 12.0$}        & {$40.0/60.0$}        & \multirow{2}{*}{{Germany}}   \\
                                                                                & HC (60)\tnote{a}  & {$57.6\pm12.2$}            & {$46.7/53.3$}      &           \\
            \multirow{2}{*}{\parbox[l]{3.2cm}{PRJDB4176 \cite{yachida2019metagenomic}} }   & CRC (258)\tnote{a}    & {$NA$}               & {$NA$}                  & \multirow{2}{*}{{Japan}}   \\
                                                                                 & HC (251)\tnote{a} & {$NA$}                     & {$NA$}              &           \\
            \multirow{2}{*}{\parbox[l]{3.2cm}{PRJEB10878 \cite{yu2017metagenomic}}}  & CRC (75)\tnote{a}     & {$65.9\pm 9.1$}       & {$40.3/59.7$}          & \multirow{2}{*}{{China}}  \\
                                                                                 & HC (53)\tnote{a}  & {$61.8\pm 9.3$}            & {$32.3/67.7$}      &           \\
            \multirow{2}{*}{total}                                           & CRC (605)         &                            &                            &       \multirow{2}{*}{}         \\
                                                                                & HC (619)          &                            &                            &            \\  \hline
        \end{tabular}%
        \begin{tablenotes}
            \footnotesize
            \item[a] Because of metadata and/or sequence-processing issues, these numbers are different from the original sample reported in the articles. NA means that some of the characteristic data in BioProject are missing, which makes it impossible to calculate the combined stats.
        \end{tablenotes}
    \end{threeparttable}
    \label{tab:sources1}%
    \vskip6pt
\end{table*}%

\subsection{Model for the binary matrix} 

\subsubsection{Introduction for the binary matrix} 

The logistic model with interaction terms based on the exhaustive method can be used to explain when there are complex interaction effects between gut microbes. First of all, only the interaction terms between two covariables, i.e., two-order interaction, are considered, as shown below
\begin{equation}
    \begin{aligned}
        \log \frac{\pi_i}{1-\pi_i} 
        = \alpha_0+ \sum_{v=1}^p \alpha_v x_{i,v} + \sum_{v=1}^p \sum_{w \neq v} \beta_{vw} x_{i,v} x_{i,w},
    \end{aligned}
    \label{eq:01}
\end{equation}
where $\pi_i = P(y_i=1|\bm{x}_i),\; i=1,2,\cdots,n$, $\alpha_0$ is the intercept term, $\alpha_v,\; v=1,2,\cdots,p$ and $\beta_{vw},\; 1\leq v \neq w \leq p$ are the coefficients of the covariables and interaction terms, respectively.

Consider the item related to the $v$-th covariable in Equation~(\ref{eq:01}),
\begin{gather*}
    (\alpha_v + \sum_{w \neq v} \beta_{vw}  x_{i,w})x_{i,v},
\end{gather*}
we can regard $(\alpha_v + \sum_{w \neq v} \beta_{vw}  x_{i,w})$ as the combined effect related to the $v$-th covariable on the response variable. Moreover, the value of this effect is changeable with the $i$-th individual, this may be why the Logistic model with interaction terms can be used to alleviate the heterogeneity problem. However, when the covariable dimension is high and the inter-relationships are complex, the number of interaction items will be too large to fit the data well. Suppose that for individual heterogeneity, we are more concerned with the directions of the effects of gut microbes on CRC in individuals, i.e., whether $(\alpha_v + \sum_{w \neq v} \beta_{vw}  x_{i,w})$ is positive or negative. Therefore, a binary variable $b_{iv} \in \{+1,-1\}$ is introduced to represent the combined positive and negative effects of covariable $x_{i,v}$ on observation $y_i$ in individual $i$. Then, the interaction term can be replaced and Equation~(\ref{eq:01}) can be modified as follows:
\begin{equation}
    \begin{aligned}
        \log \frac{\pi_i}{1-\pi_i} = a_0+ \sum_{v=1}^p a_v b_{iv} x_{i,v},
    \end{aligned}
    \label{eq:02}
\end{equation}
where $a_0$ and $a_v,\;v=1,\cdots,p,$ represents the intercept term and the coefficients of covariables, $(b_{iv})_{n \times p}$ is defined as the binary matrix $\bm{B}$, as shown in Figure~\ref{fig_lrbmat}(B).

Equation~(\ref{eq:02}) can be regarded as a Logistic model with binary matrix. $b_{iv}$ and $x_{i,v}$ are corresponding one to one. $\bm{B}$ can be viewed as modifying the covariable $x_{i,v}$ into $b_{iv} x_{i,v}$, which is referring to the viewpoint in \cite{tian2014simple} and can eliminate the effect of heterogeneity of $x_{i,v}$ to some extent. In addition, each $b_{iv}$ also contains the information about other covariables $x_{i,w}, w \neq v$, and indicates the variable interaction which is graphically proved in the following.

Furthermore, the binary matrix $\bm{B}$ can be associated with higher-order interactions. Without loss of generality, the Logistic model with both second-order and third-order interaction terms is considered as follows:
\begin{equation}
    \begin{aligned}
        \log \frac{\pi_i}{1-\pi_i}& = \alpha_0+ \sum_{v=1}^p \alpha_v x_{i,v} + \sum_{v=1}^p \sum_{w \neq v} \beta_{vw} x_{i,v} x_{i,w} \\
        &\quad \quad + \sum_{v=1}^p \sum_{w \neq v} \sum_{t \neq (w, v)} \gamma_{vwt} x_{i,v} x_{i,w} x_{i,t},
    \end{aligned}
    \label{eq:03}
\end{equation}
where $\gamma_{vwt}$ represents the coefficients of third-order interactions. Similarly, only the terms related to $x_{i,v}$ is considered, i.e.:
\begin{equation*}
    \begin{aligned}
        (\alpha_v + \sum_{w \neq v} \beta_{vw} x_{i,w}  + \sum_{w \neq v} \sum_{t \neq (w, v)} \gamma_{vwt} x_{i,t} x_{i,w})x_{i,v}.
    \end{aligned}
\end{equation*}
Then, if the coefficient of $x_{i,v}$ is conducted the process which is analogous to the analysis in Equation~(\ref{eq:01}) and Equation~(\ref{eq:02}) twice, it can be modified as follows:
\begin{equation}
    \begin{aligned}
        &(\alpha_v + \sum_{w \neq v} \beta_{vw} x_{i,w}  + \sum_{w \neq v} \sum_{t \neq (w, v)} \gamma_{vwt} x_{i,t} x_{i,w})x_{i,v} \\
        &\Longleftrightarrow  (\alpha_v^{\prime} + \sum_{w \neq v} a_w^{\prime} b_{iw}^{\prime} x_{i,w} )x_{i,v} \\
        &\Longleftrightarrow a^{\prime \prime }_v b_{iv}^{\prime \prime} x_{i,v},
    \end{aligned}
    \label{eq:05}
\end{equation}
where $b_{iw}^{\prime},b_{iv}^{\prime \prime}$ are binary variables, $a_w^{\prime},a^{\prime \prime }_v$ are their corresponding coefficients. Thus, Equation~(\ref{eq:03}) can be also modified as follows:
\begin{equation}
	\begin{aligned}
		\log \frac{\pi_i}{1-\pi_i} = a^{\prime \prime }_0+ \sum_{v=1}^p a^{\prime \prime }_v b_{iv}^{\prime \prime} x_{i,v}.
	\end{aligned}
	\label{eq:04}
\end{equation}

\begin{figure}[h]
    \centering
    \includegraphics[width=8.5cm]{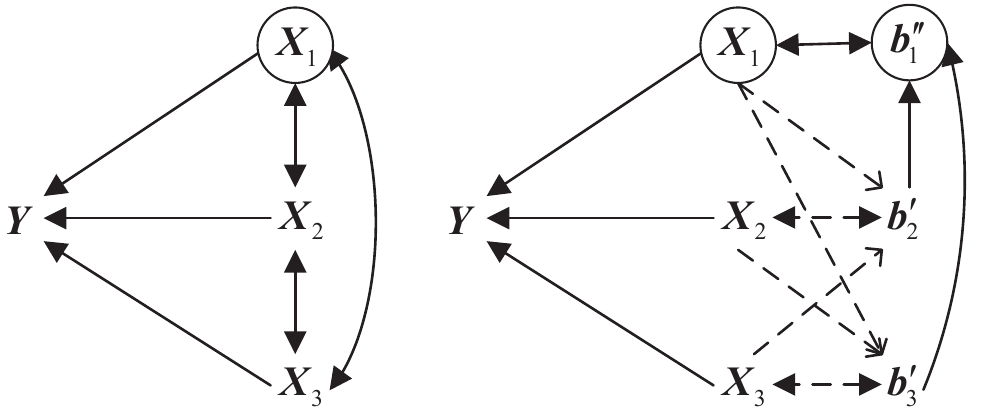}
    \caption{Graph to expatiate the binary matrix. The left and right panel represent the commonly used model with interaction terms and the proposed model, respectively. The single arrow represents the calculation process, which the dotted one and solid one represent the calculation process of the previous stages and the last stage, respectively. The double arrow represents the interaction or dot product.}
    \label{fig05}
\end{figure}

For better illustration, a graphical approach is proposed to explain why this kind of binary variable could contains the information about interaction terms of any order and relieve the influence of the individual heterogeneity. Figure~\ref{fig05} shows an example with second-order and third-order interaction terms, including the response variable $\bm{Y}$ and the covariables $\bm{X}_1,\;\bm{X}_2,\;\bm{X}_3$ with complex interaction relations. The left panel represents the model using interaction terms, and there are four terms related to $\bm{X}_1$: $\bm{X}_1$, $\bm{X}_1 \bm{X}_2,\; \bm{X}_1 \bm{X}_3$ and $\bm{X}_1 \bm{X}_2 \bm{X}_3$. The right panel is the diagram of the proposed model. When only considering the calculation of binary variable related to $\bm{X}_1$, i.e., $\bm{b}_1^{\prime \prime}$, we can find that the two binary variables $\bm{b}_2^{\prime}$ and $\bm{b}_3^{\prime}$ contain the information of second-order interaction related to $\bm{X}_2,\;\bm{X}_3$, like the first arrow in Equation~(\ref{eq:05}). Based on this, $\bm{b}_2^{\prime} \bm{X}_2 ,\; \bm{b}_3^{\prime} \bm{X}_3$ and $\bm{X}_1$ are used to calculate $\bm{b}_1^{\prime \prime}$, like the second arrow in Equation~(\ref{eq:05}), which make $\bm{b}_1^{\prime \prime}$ contain the information about second-order and third-order interaction terms associated with $\bm{X}_1$. Moreover, all binary variables in all stages are changed with the individual, which indicates they also neutralize the influence of the individual heterogeneity to some extent. In practice, the $\bm{b}_2^{\prime}$ and $\bm{b}_3^{\prime}$ are not calculated in the process, $\bm{b}_1^{\prime \prime}$ is calculated directly. 

\subsubsection{Algorithm for the binary matrix}

Since the element $b_{iv}$ of $\bm{B}$ is related to the $v$-th gut microbe and the $i$-th individual simultaneously, it is not easy to fit Equation~(\ref{eq:02}) when both of $a_v$ and $b_{iv}$ are unknown. Considering the generalization of the use of $\bm{B}$, we expect it to be estimated separately from Equation~(\ref{eq:02}). A simple method to estimate $\bm{B}$, called LRBmat, is proposed, which is referring to the approach of the effective enhancement section in \cite{tian2014simple}. In view of the relationship between $b_{iv}$ and $\alpha_v + \sum_{w \neq v} \beta_{vw} x_{i,w}$, the following points may be involved:
\begin{itemize}
    \item $b_{iv}$ is a binary variable and can be calculated by the Logistic model;
    \item $\alpha_v + \sum_{w \neq v} \beta_{vw} x_{i,w}$ essentially represents the main effect and interacted effects of $x_v$, it is expected to take into account the information about $x_v$ and interactions between $x_v$ and $x_w, w \neq v$;
    \item not all gut microbes have strong interactions, so only a few gut microbes with high correlation with $x_v$ can be considered.
\end{itemize}

To sum up, only the target $v$-th gut microbe and its most relevant gut microbes are used to estimate the value of $b_{iv}$, which reflects the combined effect of the gut microbe on each individual. For a certain covariable, denoted as $\bm{x}_{,v_{0}}$, the model to estimate the corresponding $b_{iv_0}$ is shown as follows:
\begin{equation}
    \begin{aligned}
        \log & \left(\frac{\operatorname{Pr}\left(y_{i}=1 \mid v_{0}\right)}{1-\operatorname{Pr}\left(y_{i}=1 \mid v_{0}\right)}\right) \\
        = & c+\sum_{s_1=0}^{r} \alpha_{s_1} x_{i, v_{s_1}}+\sum_{s_2 = 1}^{r} \beta_{s_2} x_{i, v_{0}} x_{i, v_{s_2}},
    \end{aligned}
    \label{eq:07}
\end{equation}
where $\text{Pr}\left(y_{i}=1 \mid v_{0}\right) $ represents the probability of $y_{i}=1$ when only uses the $v_{0}$-th and its related covariables, $c$ is the intercept term, $\alpha_{s_1}, s_1 = 0, 1, \cdots, r$ and $\beta_{s_2}, s_2 = 1,2, \cdots, r$ denote the coefficients of covariates and interaction terms, $v_{s_1}, v_{s_2} \in \{1,2,\cdots,p\}$, and $\bm{x}_{, v_{s_1}}, s_1= 1, \cdots, r$ is the covariable with the $s_1$-th highest correlation with $\bm{x}_{, v_{0}}$. The interaction terms, $x_{i, v_{0}} x_{i, v_{s_2}}$, $s_2=1, \cdots, r$, which are all related to $\bm{x}_{,v_{0}}$, are added to the model to avoid obtaining the same results of different $v_0$, when they have the same $r$ correlated covariables.

There are several things about this model that are worth elaborating on:

1. The purpose of the model. 
The aim of Equation~(\ref{eq:07}) is to use one specific gut microbe, like $\bm{x}_{,v_{0}}$, to judge the positive or negative effect of this gut microbe on each observations. In the training process with $n^{\prime}$ observations, the training result obtained for the $v_{0}$-th gut microbe is $\bm{\text{Pr}}_{n^{\prime}}= \left(\text{Pr}(y_{k_{1}}=1 \mid v_{0}), \text{Pr}(y_{k_{2}}=1 \mid v_{0}), \cdots, \text{Pr}(y_{k_{n^{\prime}}}=1 \mid v_{0})\right), k_{i} \in \{1,2, \cdots, n\} $. In the case of the lowest error, the value $\text{Pr}\left(y_{k_{i}}=1 \mid v_{0}\right)$ can be used for classification and if $y_{k_i}$ tends to be classified as $1$, indicating that the $v_{0}$-th gut microbe has a positive effect on the $k_{i}$-th observation, i.e., $b_{k_{i}, v_{0}}=1$, and otherwise, it has a negative effect or no effect, i.e., $b_{k_{i}, v_{0}}=-1$.

2. The adoption of the other $r$ gut microbes.
There are many researches of taking correlations among covariates into account in models. For example, in gene expression, \cite{hikichi2020correlation} proposed a correlation-centric approach by selecting the minimal gene sets and using a multiple logistic regression model for breast cancer prediction. Similarly, \cite{liu2017structured} developed a structured penalized logistic regression model by selecting highly correlated features in gene expression data. We draw on this idea by including the most relevant $r$ covariables of each target covariable in the proposed regularized logistic model, where the regularization can reduce overfitting of the model. At the same time, because the correlations are symmetrical, it is necessary to design these product terms $x_{i, v_{0}} x_{i, v_{j}}, j=1, \cdots, r $, which all related to the target gut microbe, i.e., the $v_{0}$-th one, to avoid overlapping of results.

The Algorithm~(\ref{alg:02}) shows the detailed process of the binary matrix construction. For a new observation, $x^{*}$, its positive or negative effect of the $v$-th gut microbe can be calculated based on the threshold $p_{0 v}$ for the corresponding logistic model trained in Algorithm~(\ref{alg:02}), and then the corresponding binary value can be obtained, i.e.,  $b_{x^{*} v}, v=1, \cdots, p $.
\begin{algorithm}[h]  
    \caption{ Construct a binary matrix.}
    \label{alg:02}
    \begin{algorithmic}[1]
        \REQUIRE Data matrix $\bm{X}_{n \times p}$; The labels of $n$ samples $\bm{y}$;
        \ENSURE Binary matrix $\bm{B}$; Effective matrix $\bm{Emat}$;
        \STATE Calculate the correlation matrix $\bm{Cor}_{X}$ of data $\bm{X}_{n \times p}$.
        \FOR{each $v_0\;\text{in}\;\{1, 2, \cdots, p\}$}
        \STATE Select the $r$ covariables, $\bm{x}_{,v_{1}}, \cdots, \bm{x}_{,v_{r}}$, with the highest correlation with $\bm{x}_{,v_{0}}$ and define $\bm{X}_{n \times (r+1)}=\left(\bm{x}_{,v_{0}}, \bm{x}_{,v_{1}}, \cdots, \bm{x}_{,v_{r}}\right)$;
        \STATE Bring $\bm{X}_{n \times (r+1)}$ into Equation~(\ref{eq:07}) for training and the result is $\bm{p}_{v_0}=(p_{1 v_0}, \cdots, p_{n v_0})^{T}$;
        \STATE Determine the optimal threshold $p_{0 v_0}$ for $\bm{p}_{v_0}$ such that the logistic model has a minimum prediction error;
        \STATE Obtain the binary vector $\bm{b}_{v_0}=(b_{1 v_0}, \cdots, b_{n v_0})^{T}$, where $b_{i v_0}={1}\; \text{when}\; p_{i v_0}>p_{0 v_0}\; \text{and} \; b_{i v_0}={-1}\; \text{otherwise},\;i = 1,2,\cdots,n$;
        \ENDFOR
        \STATE Save the binary matrix, $\bm{B} =\left(\bm{b}_{1}, \bm{b}_{2}, \cdots,\bm{b}_{p}\right)$;\\
        Calculate the effective matrix, $\bm{Emat} = \bm{B} \ast \bm{X}$, where $\ast$ denotes Hadamard product;\\
        \RETURN $\bm{B}, \bm{Emat}$.
    \end{algorithmic}
\end{algorithm}

As mentioned before, the binary matrix $\bm{B}$ containing the information about gut microbial interactions and individual heterogeneity is inferred separately from Equation~(\ref{eq:02}). Therefore, its usage has very strong generalization, which means that $\bm{B}$ can be combined with any classification method to produce better results, as shown in Figure~\ref{fig_lrbmat}(C). This will be validated detailedly in the experiments in Section~\ref{s:result}.

\subsection{Classification and association rule mining of CRC based on the binary matrix} 

The effective matrix, $\bm{Emat} = \bm{B} \ast \bm{X}$, obtained from Algorithm~\ref{alg:02}, can be regarded to add information about the combined effect of the gut microbes on each individual for the original data $\bm{X}$. Therefore, in order to verify the superiority of $\bm{Emat}$ in the classification of CRC, four commonly used machine learning models, such as SVM, Logistic, RF and XGBoost, are adopted, and the following four evaluation metrics are used to evaluate the qualities of classifiers: True Positive Rate $\left(TPR\right)=\frac{TP}{TP+FN}$, False Positive Rate $\left(FPR\right)=\frac{FP}{FP+TN}$, Accuracy $\left(ACC\right)=\frac{TP+FP}{TP+FN+FP+TN}$ and Area Under Curve $\left(AUC\right)$.

Furthermore, the binary matrix $\bm{B}$, which reflects the effect relationship between several microorganisms and individuals, can also be used to detect the associations among two or more microorganisms on CRC and HC. Thus, the association rule mining algorithm \cite{luo2021computational,liu2021maniea} is adopted which can detect the most likely rules and significantly reduce the workload. The process is illustrated in Figure~\ref{fig_lrbmat}(D).

An association rule is shown as $\{po_{a},\cdots, ne_{b},\cdots\}\Rightarrow \{ \text{CRC or HC}\} $, where $po_{a}$ and $ne_{b}$ denote the positive and negative effects of $a$-$th$ and $b$-$th$ gut microorganism on individuals, respectively. We use left-hand-side $(LHS)$ to denote $\{po_{a},\cdots, ne_{b},\cdots\}$ and right-hand-side $(RHS)$ to denote $\{\text{CRC or HC}\}$ for an association rule. Four metrics, support $\left(supp\right)$, confidence$\left(conf\right)$, lift $\left(lift\right)$ and improvement $\left(ipv\right)$, will be used to detect possible combinations. Firstly, fixed $supp_{min} = 0.01,\;conf_{min}=0.8$, where $supp$ indicates the probability of the gut microbial combination occurring in the sample, and $conf$ indicates the probability of the sample being CRC given the occurrence of the combination. These two metrics can be used to filter out combinations that occur too infrequently or are not associated with CRC. Secondly, the $lift$ and $ipv$ are used to filter useful rules, with $lift>1$ or $ipv>0$ indicating that the rule is helpful for detecting CRC.

The formulas for these metrics are as follows:
\begin{equation}
    \label{eq:05}
    \begin{split}
        \operatorname{supp}(LHS)&=\frac{\operatorname{occur}(LHS)}{n}(\times 100 \%), \\
        \operatorname{conf}(LHS\Rightarrow RHS)&=\frac{\operatorname{supp}(LHS \bigcup RHS)}{\operatorname{supp}(LHS)}\\ &=P(RHS \mid LHS),\\
        \operatorname{lift}(LHS\Rightarrow RHS)&=\frac{\operatorname{conf}(LHS\Rightarrow RHS)}{\operatorname{supp}(RHS)}\\ &=\frac{\operatorname{supp}(LHS \Rightarrow RHS)}{\operatorname{supp}(LHS) * \operatorname{supp}(RHS)},\\
        \text {ipv}(LHS \Rightarrow RHS)&=\min _{LHS^{\prime} \subset LHS} \left[\right. \operatorname{conf} (LHS \Rightarrow RHS)\\ &-\operatorname{conf}\left(LHS^{\prime} \Rightarrow RHS\right) \left.\right],  \\
    \end{split}
\end{equation}
where $\operatorname{occur}(LHS)$ is the number of times the gut microbial combination of $LHS$ appears in $\bm{B}$ and $n$ is the total number of samples.

\section{Results} 
\label{s:result}

\subsection{Gut microbial screening}

It is unlikely that all gut microorganisms in the microbial community are closely related with CRC \cite{bang2019establishment}, and using classifiers directly to the high-dimensional data is less effective, thus, it's necessary to screen these gut microorganisms. The test for contingency table, a simple and common variable selection method, is adopted \cite{tuyl2008inference,sulewski2019some}. Specifically, based on a $2\times 2$ contingency table, each gut microorganism is screened by a Chi-square test, or a Fishers exact test when the conditions of the Chi-square test are not met. The screening algorithm with the idea of cross-validation is displayed in Section~S1 of the supplementary material.

After 100 times of 5-fold cross-validation screening, 159 species of gut microorganisms are finally screened out. Table~\ref{tab:selection1} shows these microorganisms at different levels (Kingdom, Phylum and Class).
\begin{table*}[!htbp]
    \centering
    \vspace*{-12pt}
    \caption{Changes after and before selection of gut microbes at different levels.}
    \begin{tabular}{@{}ccc@{}}
        \hline
        Level   & \parbox[c]{3cm}{\centering{No. of level groups \\after (n) / before (n)}}  & \parbox[*]{12cm}{Level groups \\ No. of species after (n) $/$ before (n)}              \\ \hline \rule{0pt}{10pt}
        Kingdom & 3/4            & \parbox[l]{12cm}{Archaea (2/4)\tnote{1}, Bacteria (155/715), Viruses (2/117)}                                       \\  \rule{0pt}{22pt}
        Phylum  & 10/18          & \parbox[l]{12cm}{Euryarchaeota (2/4), Actinobacteria (18/98), Bacteroidetes (27/112), \\Candidatus Saccharibacteria (1/2), Firmicutes (77/332), Fusobacteria (8/13), \\Proteobacteria (20/135), Spirochaetes (2/8), Synergistetes (2/5), Viruses (2/117)} \\  \rule{0pt}{35pt}
        Class   & 16/28         & \parbox[l]{12cm}{Methanobacteria (2/4), Actinobacteria (18/98), Bacteroidia (27/107), \\Candidatus Saccharibacteria (1/2), Bacilli (15/134), Clostridia (50/139), \\Erysipelotrichia (6/23), Negativicutes (6/36), Fusobacteriia(8/13), \\Betaproteobacteria (2/28), Deltaproteobacteria (2/8), Epsilonproteobacteria (2/15), \\Gammaproteobacteria (14/80), Spirochaetia (2/8), Synergistia (2/5), Viruses (2/117)}\\ \hline
    \end{tabular}
    \label{tab:selection1}
\end{table*}

The results of microbial screening are meaningful and can be verified by many literatures. At the kingdom level, the main gut microorganisms screened are Archaea, Bacteria, and Viruses. \cite{coker2020altered} mentioned that methanogenic archaea was significantly reduced in fecal samples of CRC patients, corresponding that two species of methanobrevibacter in Archaea are screened out. \cite{zou2018dysbiosis} found that the intestinal flora plays a critical role in gut microbiota, confirming that most of the screened gut microorganisms are Bacteria $(155/159)$. At the phylum level, there are mainly Firmicutes, Bacteroidetes, Proteobacteria and Actinobacteria. \cite{guo2021ginger} mentioned the dominance of the above-mentioned intestinal flora in healthy adults. In additon, \cite{janney2020host} summarized some bacteria that might cause dysbiosis in CRC patients, such as \emph{F. nucleatum}, \emph{B. fragilis} and other bacteria, which are all included in the screening results. In summary, the screened gut microorganisms are all significant and related to CRC.

\subsection{Hypothesis checking for individual heterogeneity}

The individual heterogeneity is the main concern in the paper, thus, its existence in the real data needs to be examined. Here, two simple methods are applied after microbial selection.

The first is to use a two-sample $t$-test \cite{zhu2019two,cao2018two}. The main step is to test whether there is a significant difference between the mean values of one microorganism in CRC and HC populations. If the result is not significant, it seems to contradict the findings that this microorganism is screened out and considered to be related with CRC. Therefore, it's reasonable to assume that this microorganism may have different effects on different individuals, i.e., the individual heterogeneity. By the two-sample $t$-test, 83 of the 159 kinds of gut microorganisms are considered to satisfy the hypothesis.

The other is to use the coefficient of logistic model, which can indicate the direction of impact of each gut microbe on CRC. The samples are randomly divided into five parts and logistic models are trained respectively. If there is heterogeneity in the effect of one microbe on individuals, then, coefficients of this microbe in the five models may appear to be different directions, positive or non-positive. After repeating the above steps 10 times, among 159 gut microorganisms, there are 53 microorganisms that have inconsistent coefficient directions more than 5 times.

In both of the two methods, 33 kinds of gut microorganisms are considered as having heterogeneity problem. Table~\ref{tab:two tests} shows the numbers of the 33 gut microorganisms at different levels. Some evidence for the results can be found in existing literatures. For example, two species of \emph{Lactobacillus} satisfy the hypothesis, which is consistent with the findings in \cite{wang2019lactobacillus}. Similarly, \cite{jia2020metagenomic} mentioned that the bile acid metabolism would have beneficial and harmful effects on host health, and the sources of this microbial bile salt hydrolase were mainly the Phylums of Firmicutes and Actinobacteria, which validate that there are 22 of the 33 microbes belonging to these two Phylums. Although the hypothesis testing may not be necessarily comprehensive, the results above still have certain reliability and validity, which indicate the rationality of the heterogeneity to some extent.
\begin{table}[!htbp]   
    \centering
    \vspace*{-12pt}
    \caption{The number of gut microbes at different levels after two hypothesis checking methods.}
    \begin{tabular}{@{}ccl@{}}
        \hline
        {\parbox[c]{0.5cm}{\centering{Level}}}   & {\parbox[c]{1cm}{\centering{No. of level groups \\(n)}}} & {\parbox[*]{5.5cm}{Level groups \\ No. of species (n)}}   \\ \hline \rule{0pt}{10pt}
        {Kingdom} & {1}     & \parbox[l]{5.5cm}{Bacteria (33)}                            \\ \rule{0pt}{22pt}
        {Phylum}  & {5}     & \parbox[l]{5.5cm}{Actinobacteria (2), Bacteroidetes (8), \\Firmicutes (20), Fusobacteria (2), \\Proteobacteria (1)} \\  \rule{0pt}{34pt}
        {Class}   & {8}     & \parbox[l]{5.5cm}{Actinobacteria (2), Bacteroidia (8), \\Bacilli (3), Clostridia (10), \\Erysipelotrichia (4), Negativicutes (3), \\Fusobacteriia (2), \\Gammaproteobacteria (1)} \\ \rule{0pt}{34pt}
        {Order}   & {8}     & \parbox[l]{5.5cm}{Actinomycetales(2), Bacteroidales(8), \\Lactobacillales(3), Clostridiales(10), \\Erysipelotrichales(4), \\Selenomonadales(3), \\Fusobacteriales(2), Aeromonadales (1)} \\ \hline
    \end{tabular}%
    \label{tab:two tests}%
\end{table}

\subsection{Classification of colorectal cancer}

\begin{table}[!htbp] 
	\centering
	\vspace*{-6pt}
	\caption{Classification result for real data.}
	\setlength{\tabcolsep}{0.6mm}{
		\begin{tabular}{@{}lcccc@{}}
			\hline
			\multicolumn{5}{c}{$\bm{Emat}$}                                                     \\ \hline
			 {classifiers}  & {TPR}        & {FPR}       & {ACC}       & {AUC}       \\ \hline
			 {SVM}          & {0.8731}  & {0.3639} & {0.6962} & $\bm{0.7810}$ \\
			 {logistic}     & {0.7656}  & {0.3219} & {0.7096} & $\bm{0.7810}$ \\
			 {RF}           & {0.7443}  & {0.2829} & {0.7277} & {0.8056} \\
			 {XGBoost}      & {0.7629}  & {0.2908} & {0.7282} & $\bm{0.8069}$ \\ \hline
			\multicolumn{5}{c}{$\bm{X}$}                                                      \\ \hline
			 {classifiers}  & {TPR}        & {FPR}       & {ACC}       & {AUC}       \\ \hline
			 {SVM}          & {0.7798} & {0.4053} & {0.6262} & {0.7352} \\
			 {logistic}     & {0.6360} & {0.3631} & {0.5914} & {0.6713} \\
			 {RF}           & {0.7175} & {0.2589} & {0.7269} & $\bm{0.8061}$ \\
			 {XGBoost}      & {0.7418} & {0.2875} & {0.7242} & {0.8001} \\ \hline
	\end{tabular} }
	\vspace*{-6pt}
	\label{tab:Classification result 2}
\end{table}

In this section, LRBmat and $\bm{Emat}$ will be proved to enhance the classification performance of machine learning methods on the real data for CRC.

Table~\ref{tab:Classification result 2} shows the average classification results after $20$ times of 5-fold cross-validation based on the screened gut microbial data $\bm{X}$. The top and bottom panels correspond to $\bm{Emat}$ and $\bm{X}$, respectively. We can easily find that the results of $\bm{Emat}$ are better than those of $\bm{X}$ in all four classifiers, which validate the effectiveness of LRBmat and show that the introduction of LRBmat can improve the diagnosis of CRC disease. The main reasons may be as follows:

1. LRBmat takes more account of the characteristics of the gut microbial data, such as the intrinsically microbial correlations, which brings more useful information. At the same time, LRBmat modifies the original gut microbial data, which maybe avoid or weaken the influence of heterogeneity.

2. LRBmat contains only $1$ and $-1$, and the gut microbial data are highly skewed with most of the values closing to 0, thus, the combination of LRBmat and $\bm{X}$, i.e., $\bm{Emat}$, allows the differences among the data to be amplified and makes classification more favourable.

Furthermore, we also performed two kinds of simulations based on generated datasets with complex interactions and individual heterogeneity. The first one compares the proposed LRBmat method with the logistic model with different order interaction terms, which demonstrated the superiority of LRBmat, and we also proved this conclusion theoretically. The another one is consistent with the real data analysis above, indicating that the binary matrix can enhance the classification effect of machine learning methods. The detailed analyses are shown in Section~S2 and S3 of the supplementary material.

\subsection{Association rule mining}

\begin{figure}[ht]
	\centering
	\vspace*{-6pt}
	\includegraphics[scale=0.18]{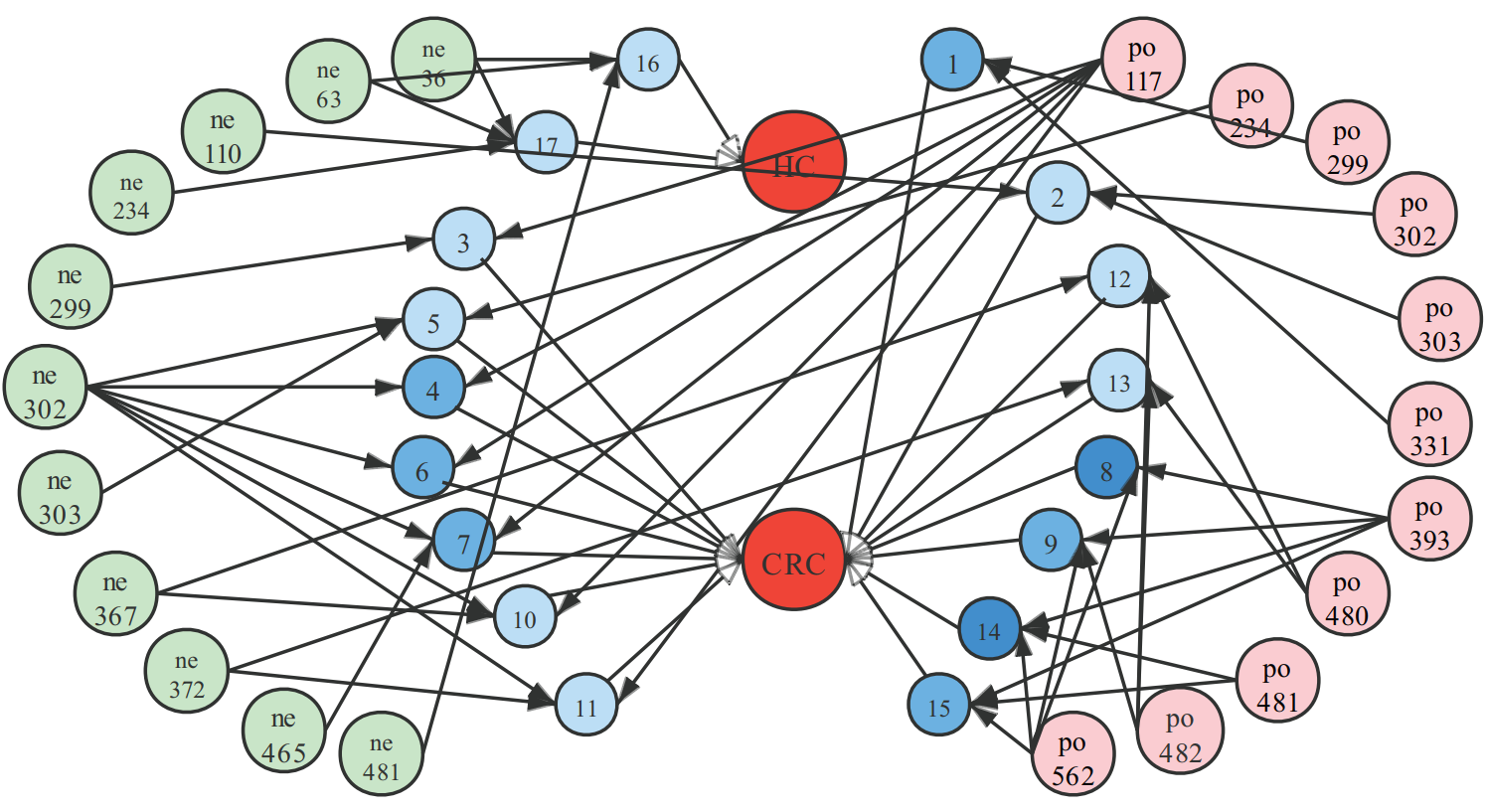}
	\caption{Illustration of association rules. Blue circles indicate rules, with darker colours indicating higher $supp$; light red and light green circles indicate positive and negative effects of different microorganisms, respectively; solid and empty arrows represent $LHS, RHS$, respectively.}
	\label{fig04}
\end{figure}

After $5$ times association rule mining, the association rules occurred more than $3$ times are retained. $452911$ rules related to CRC and $355$ rules related to HC are obtained, respectively. Some representative rules are selected and shown in Table~\ref{tab:Rule} in three parts, which are graphically illustrated in Figure~\ref{fig04}. The detailed analysis is as follows:
\begin{enumerate}
	\item \textbf{Positive and negative effects in association rules.}
	In the first part of Table~\ref{tab:Rule}, association rules show different effects of the same gut microbe on different individuals, i.e., individual heterogeneity. Lactobacillus (299, 302, 303) in rules 1-5 show this phenomenon, which is consistent with the findings of \cite{wang2019lactobacillus} mentioned in Section ~\ref{sec:introduction}. Another phenomenon is that the same rule may contain both positive and negative effects, such as rules 2-7 and 10-13, which indicates the complex interactions.
	On the other hand, \cite{huang2019reasonable} suggested some gut microbes favour the carcinogenic effect of CRC, such as ETBF (177), \emph{P. anaerobius} (480) and \emph{F. nucleatum} (562). These microbes are essentially shown to have positive effects in the detected rules.
	We also find that the gut microbes in one association rule may all come from the same phylum. For example, the gut microbes in rules 1,5 and 12-13 all belong to the phylum of Firmicutes, and \cite{flemer2017tumour} confirmed that the Firmicutes were releted to CRC.
	\item \textbf{Extension of association rules.}
	The second part of Table~\ref{tab:Rule} shows two rules and the corresponding extended rules. For example, the combined effect of \emph{Parvimonas micra} (393) and \emph{F. nucleatum} (262) is shown in rules 8 and 9, by contrast, rule 9 has an additional positive effect of \emph{Peptostreptococcus unclassified} (482). Thus, rule 9 can be regarded as an extension of rule 8. In this case, if rule 8 is present in some individuals, then we have enough incentive to further determine if rule 9 is also present, rather than aimlessly searching for other possible rules. Therefore, the extended association rule might not only reduce the workload of further research, but also perhaps validate the existing combined effect between multiple microbes on CRC.
	\item \textbf{Similarity of association rules.}
	The third part of Table~\ref{tab:Rule} shows some similar rules. For example, the only differences between rule 10 and 11 is the gut microbes 367 and 372, which belong to the same Genus Clostridium, so the two rules are similar in a sense. Rules 16 and 17 can also be described as similar if considered at the phylum level, where microbes 234 and 481 belong to Phylum Firmicutes. Similarity rules can merge the relationship between microbes to that between genus, or phylum, etc., further illustrating the validity of our mined association rules. In addition, the mined rules combined with known genus or phylum related to CRC, can guide us to mine new association rules.
\end{enumerate}

\begin{table}[H] 
	\centering
	\vspace*{-6pt}
	\caption{Some of the association rules}
	\begin{threeparttable}
			\setlength{\tabcolsep}{0.6mm}{
				\begin{tabular}{@{}lcccccc@{}}
					\hline
					No.    & $lhs$                              & $rhs$  & $supp$ & $conf$ & $lift$ & $ipv$      \\ \hline
					1      & {$po_{299}$,$po_{331}\tnote{a}$}            & {CRC}  & 0.122  & 0.842  & 1.703  & 0.024  \\  
					2      & {$po_{302}$,$po_{303}$,$ne_{110}$} & {CRC}  & 0.054  & 0.805  & 1.628  & 0.805  \\
					3      & {$po_{117}$,$ne_{299}$}            & {CRC}  & 0.072  & 0.846  & 1.712  & 0.035  \\
					4      & {$po_{117}$,$ne_{302}$}            & {CRC}  & 0.120  & 0.831  & 1.680  & 0.019  \\
					5      & {$po_{234}$,$ne_{302}$,$ne_{303}$} & {CRC}  & 0.077  & 0.969  & 1.961  & 0.011  \\
					&                                    &        &        &        &        &        \\
					6      & {$po_{117}$,$ne_{302}$}            & {CRC}  & 0.120  & 0.831  & 1.680  & 0.019  \\
					7      & {$po_{117}$,$ne_{302}$,$ne_{465}$} & {CRC}  & 0.116  & 0.840  & 1.700  & 0.003  \\
					8      & {$po_{393}$,$po_{562}$}            & {CRC}  & 0.160  & 0.916  & 1.853  & 0.003  \\
					9      & {$po_{393}$,$po_{482}$,$po_{562}$} & {CRC}  & 0.118  & 0.923  & 1.868  & 0.002  \\
					&                                    &        &        &        &        &        \\
					10     & {$po_{117}$,$ne_{302}$,$ne_{367}$} & {CRC}  & 0.067  & 0.845  & 1.710  & 0.015  \\
					11     & {$po_{117}$,$ne_{302}$,$ne_{372}$} & {CRC}  & 0.069  & 0.859  & 1.737  & 0.021  \\
					12     & {$po_{480}$,$po_{482}$,$ne_{367}$} & {CRC}  & 0.062  & 0.938  & 1.898  & 0.008  \\
					13     & {$po_{480}$,$po_{482}$,$ne_{372}$} & {CRC}  & 0.065  & 0.963  & 1.949  & 0.033  \\
					14     & {$po_{393}$,$po_{481}$,$po_{562}$} & {CRC}  & 0.146  & 0.918  & 1.857  & 0.000  \\
					15     & {$po_{393}$,$po_{482}$,$po_{562}$} & {CRC}  & 0.118  & 0.923  & 1.868  & 0.002  \\
					16     & {$ne_{36}$,$ne_{63}$,$ne_{481}$}   & {HC}   & 0.067  & 0.804  & 1.590  & 0.804  \\
					17     & {$ne_{36}$,$ne_{63}$,$ne_{234}$}   & {HC}   & 0.067  & 0.804  & 1.590  & 0.804  \\  \hline
			\end{tabular}}
			\label{tab:Rule}%
			\begin{tablenotes}
				\footnotesize
				\item[a] At the phylum level, gut microorganisms and corresponding numerical subscripts are as follows: Actinobacteria (36, 63), Bacteroidetes (110, 117), Firmicutes (234, 299, 302, 303, 331, 367, 372, 393, 465, 480, 481, 482), Proteobacteria (582, 584, 585, 590), Fusobacteria (562). At the species level, some specific gut microorganisms and corresponding numerical subscripts are Lactobacillus (299, 302, 303), \emph{ETBF} (177), \emph{P. anaerobius} (480) and \emph{F. nucleatum} (562).
			\end{tablenotes}
	\end{threeparttable}
	\vskip-6pt
\end{table}
	
\section{Conclusion}
\label{s:conclusion}

The effective diagnostic classification and etiological analysis of cancers like CRC are still worthy of long-term research. This study proposes a simple method based on Logistic regression, called LRBmat, which not only contains information about the gut microbial interactions with any order, but also can reduce or avoid the effect of individual heterogeneity, thus improving the classification effect of CRC. Moreover, LRBmat has a powerful generalization. In addition to its application in CRC, LRBmat can be used in similar data with covariable interactions and individual heterogeneity, for example, genes or MicroRNAs data involved in the pathogenesis of complex diseases \cite{huang2022updated,zhou2021multi,bofill2021tumor}, and cross-subsets data in disease treatment \cite{roider2020dissecting,crinier2018high}. It also can greatly improve the classification ability of Logistic model, and more importantly, LRBmat can combine with any machine learning method and enhance them to a certain extent. Both simulation studies and a real data analysis display the high power of our proposed method. Moreover, the binary matrix can also be used to detect the complex combined effect of multiple gut microbes on CRC, through association rule algorithm. The detected association rules are verified in the existing literatures, which provides some reference for the occurrence and development of CRC.

However, it is worth noting that the improvement of LRBmat may not be very obvious sometimes. For example, the results in real data analysis for RF and XGboost only improved 1\%. This probably because the gut microbial data is too sparse, the proportion of zeros is 84.2\%, much higher than 72.2\% in \cite{zhang2019scalable}, which leads to great heterogeneity \cite{chen2016two}. Even if the binary matrix can amplify the differences in microbial data, the effect of LRBmat will be greatly affected in this case.

There are surely many directions to extent our method, such as the algorithm of LRBmat. A more flexible $r$ is needed, since the number of correlated gut microbes for each microbe is significantly different. Alternatively, more complex models can be used to compute LRBmat, such as the Logistic Normal Polynomial (LNM) model \cite{xia2013logistic,zhang2019scalable}. In addition, LRBmat in Algorithm~(\ref{alg:02}) is calculated only once, which may not contain sufficient information about gut microbial data. Bayesian idea can be considered, such as by classifying multiple times based on sampling and then gradually optimising the binary matrix. In terms of association rules, to avoid partly overlapping problems (such as association rule expansion and similarity mentioned above), on the one hand, more advanced algorithms and tools can be selected for research, such as SAM \cite{kim2017construction} and selective association rule generation \cite{hahsler2008selective}. On the other hand, the screening of association rules can be done at a higher level to avoid excessive redundant and similar rules at the species level.

In summary, our main goal is to show the effectiveness of the proposed LRBmat and its generalization for combining with machine learning methods, and our results do show this.

\section*{Declaration of competing interest}

The authors declare that they have no known competing financial interests or personal relationships that could have appeared to influence the work reported in this paper.

\section*{Acknowledgments}

This work is supported by the Fundamental Research Funds for the Central Universities of China [531118010643].




\bibliographystyle{elsarticle-num}
\bibliography{manuscript_bib} 



\end{document}